\documentclass[12pt]{article}
\usepackage{amsthm,amsmath,latexsym,amssymb,amsfonts,amscd}
\usepackage{graphics,lscape,fancyhdr,array,stmaryrd,euscript,wrapfig}
\usepackage{empheq,slashed}
\usepackage{verbatim,slashed}
\numberwithin{equation}{section}
\usepackage{hyperref,setspace}
\usepackage{tikz-cd}
\usepackage{mathrsfs}
\usepackage[numbers,sort&compress]{natbib}
\usepackage[nottoc]{tocbibind}

\newcommand{\pl}{\partial}
\newcommand{\plb}{\bar{\partial}}

\newcommand{\be}{\begin{align}}
\newcommand{\ee}{\end{align}}

\newcommand{\aA}{{\ensuremath{\mathcal{A}}}}
\newcommand{\aB}{{\ensuremath{\mathcal{B}}}}

\newcommand{\bry}{{{\bar{y}}}}

\newcommand{\hs}{{\mathfrak{hs}}}

\newcommand{\fud}[2]{{}^{#1}{}_{#2}\,}

\newcommand{\hhbar}{{\lambda}}

\DeclareMathOperator{\sign}{sign}

\newcommand{\besubeqs}{\begin{subequations}}
\newcommand{\esubeqs}{\end{subequations}}

\usepackage{tensor}
\usepackage{todonotes}

\renewcommand{\bar}[1]{\overline{#1}}

\usepackage[all,cmtip]{xy}

\begin{document}
\pagenumbering{gobble}
\hfill
\vskip 0.01\textheight
\begin{center}
{\Large\bfseries 
Chiral Higher Spin Gravity in (A)dS${}_4$ and \\ [3mm] secrets of Chern--Simons Matter Theories}

\vspace{0.4cm}

\vskip 0.03\textheight
\renewcommand{\thefootnote}{\fnsymbol{footnote}}
Alexey \textsc{Sharapov}${}^{a}$ and 
Evgeny \textsc{Skvortsov}\footnote{Research Associate of the Fund for Scientific Research -- FNRS, Belgium}${}^{b,c}$
\renewcommand{\thefootnote}{\arabic{footnote}}
\vskip 0.03\textheight

{\em ${}^{a}$Physics Faculty, Tomsk State University, \\Lenin ave. 36, Tomsk 634050, Russia}\\
\vspace*{5pt}
{\em ${}^{b}$ Service de Physique de l'Univers, Champs et Gravitation, \\ Universit\'e de Mons, 20 place du Parc, 7000 Mons, 
Belgium}\\
\vspace*{5pt}
{\em ${}^{c}$ Lebedev Institute of Physics, \\
Leninsky ave. 53, 119991 Moscow, Russia}\\

\end{center}

\vskip 0.02\textheight

\begin{abstract}
Chiral Higher Spin Gravity with cosmological constant is constructed as a Free Differential Algebra, i.e. at the level of equations of motion, which is a smooth deformation of its flat space cousin \href{https://arxiv.org/abs/2205.07794}{arXiv:2205.07794}. Chiral Higher Spin Gravity is a unique class of local higher spin theories; its very existence implies that there is a closed and, most likely, integrable sub-sector of Chern--Simons Matter Theories, which has important consequences both for the theories themselves and for three-dimensional bosonization duality.  
\end{abstract}

\newpage
\section{Introduction}
\label{sec:}
\pagenumbering{arabic}
\setcounter{page}{2}
Higher Spin Gravities (HiSGRA) are theories \cite{Bekaert:2022poo} that synthesize several fruitful ideas in quest of solving the quantum gravity problem: (a) higher spin states are very likely to be present in any viable model of quantum gravity as suggested by (super)string theory and by AdS/CFT correspondence \cite{Maldacena:1997re,Gubser:1998bc,Witten:1998qj}; (b) extensions of gravity with higher symmetries, e.g. supergravities, should have better quantum behaviour; (c) in the high energy regime, where the perturbative quantum gravity problems are coming from, it should be possible to neglect masses. All together (a,b,c) suggest it can be instructive to look for extensions of gravity with massless higher spin fields, whose dynamics are controlled by infinite-dimensional gauge symmetries. It comes as no surprise  that HiSGRA are not easy to construct and face numerous problems already at the classical level since the masslessness makes them sensitive to genuine quantum UV problems.

There is a handful of classes of HiSGRA's that have been constructed so far, all of which are quite peculiar. In $3d$ there is one class with a plenty of topological higher spin theories with massless, partially-massless and conformal higher spin fields \cite{Blencowe:1988gj,Bergshoeff:1989ns,Campoleoni:2010zq,Henneaux:2010xg,Pope:1989vj,Fradkin:1989xt,Grigoriev:2019xmp}. In $4d$ (and all even dimensions) it is possible to construct conformal HiSGRA \cite{Segal:2002gd,Tseytlin:2002gz,Bekaert:2010ky} that is a higher spin extension of (conformal) Weyl gravity. As far as HiSGRA with propagating massless fields are concerned, there is a unique class of such theories in $4d$ -- Chiral HiSGRA \cite{Metsaev:1991mt,Metsaev:1991nb,Ponomarev:2016lrm,Skvortsov:2018jea,Skvortsov:2020wtf} and its contractions \cite{Ponomarev:2017nrr,Krasnov:2021nsq}. Chiral HiSGRA is closely related to self-dual theories \cite{Ponomarev:2017nrr}. Its contractions can be understood as higher spin extensions of SDYM and SDGR \cite{Krasnov:2021nsq}.

Originally, Chiral Theory was constructed in the light-cone gauge and in flat space \cite{Metsaev:1991mt,Metsaev:1991nb,Ponomarev:2016lrm}. Thanks to a deep interrelation between the light-cone gauge and spinor-helicity formalism \cite{Chalmers:1998jb,Chakrabarti:2005ny,Chakrabarti:2006mb,Bengtsson:2016jfk,Ponomarev:2016cwi} Chiral Theory can easily be defined without having to make explicit reference to the light-cone gauge. As is well-known, for any triplet of helicities $\lambda_{1,2,3}$ such that $\lambda_1+\lambda_2+\lambda_3>0$ there is a unique light-cone vertex and the corresponding three-point amplitude \cite{Bengtsson:1986kh,Benincasa:2011pg}: 
\begin{align}\label{genericV}
   V_{\lambda_1,\lambda_2,\lambda_3}\Big|_{\text{on-shell}} \sim 
    [12]^{\lambda_1+\lambda_2-\lambda_3}[23]^{\lambda_2+\lambda_3-\lambda_1}[13]^{\lambda_1+\lambda_3-\lambda_2}\,.
\end{align}
Chiral Theory is a unique local Lorentz invariant theory that contains at least one higher spin field with a nontrivial self-interaction. This simple input forces one to introduce massless fields of all spins (at least even), thereby, adding graviton and a scalar field. The coupling constants are uniquely fixed to be 
\begin{align}\label{eq:magicalcoupling}
    V_{\text{Chiral}}&= \sum_{\lambda_1,\lambda_2,\lambda_3}  C_{\lambda_1,\lambda_2,\lambda_3}V_{\lambda_1,\lambda_2,\lambda_3}\,, && C_{\lambda_1,\lambda_2,\lambda_3}=\frac{\kappa\,(l_p)^{\lambda_1+\lambda_2+\lambda_3-1}}{\Gamma(\lambda_1+\lambda_2+\lambda_3)}\,,
\end{align}
where $l_p$ is a constant of dimension length and $\kappa$ is an arbitrary dimensionless constant. 

Chiral Theory was shown to be one-loop finite in flat space \cite{Skvortsov:2018jea,Skvortsov:2020wtf,Skvortsov:2020gpn} directly in the light-cone gauge. Its extension to $AdS_4$ was envisaged in \cite{Metsaev:2018xip,Skvortsov:2018uru}, again in the light-cone gauge. Nevertheless, it had remained unclear if Chiral Theory admits a manifestly Lorentz-invariant formulation before it was shown in \cite{Krasnov:2021nsq} that contractions of Chiral Theory, which can be understood as higher spin extensions of SDYM and SDGR, do have a manifestly Lorentz-invariant formulation both in flat and $(A)dS_4$ spaces. In addition, contrary to the old folklore, it was shown that the interactions have a smooth deformation to (anti)-de Sitter space. 

A manifestly Lorentz-invariant formulation of Chiral Theory in flat space was found in \cite{Skvortsov:2022syz,Sharapov:2022faa} at the level of equations of motion. The equations were constructed in the form of a Free Differential Algebra \cite{Sullivan77, vanNieuwenhuizen:1982zf,DAuria:1980cmy}, thereby, fulfilling the old ideas \cite{Vasiliev:1988sa}. An important new feature of the FDA language \cite{Barnich:2010sw,Grigoriev:2012xg,Grigoriev:2019ojp,Grigoriev:2020lzu} is that it can be understood as the minimal model of the BV-BRST formulation of a given theory and, hence, contains the same information as local BRST cohomology \cite{Brandt:1997iu,Brandt:1996mh,Barnich:1994db,Barnich:1994mt,Kaparulin:2011xy}, e.g. one can study counterterms and anomalies. This is also a very useful first step towards a Lorentz-invariant action. 

HiSGRA in (anti)-de Sitter background are, on the first sight, very well supported by AdS/CFT correspondence, where HiSGRA should be duals of free/critical (Chern--Simons) vector models \cite{Klebanov:2002ja,Sezgin:2003pt,Leigh:2003gk,Giombi:2011kc}. On the second thought, however, the duality itself implies that HiSGRA must be too nonlocal for standard field theory methods to be applicable and for the theories themselves to exist in any sensible way as field theories \cite{Bekaert:2015tva,Maldacena:2015iua,Sleight:2017pcz,Ponomarev:2017qab}. At present, there is no construction of AdS/CFT duals of (Chern--Simons) vector models that makes concrete, systematic, and meaningful predictions for bulk interactions and allows one to compute holographic correlation functions.\footnote{\label{vasilkaput}The first steps \cite{Giombi:2009wh,Giombi:2010vg}, while generally positive, revealed some puzzles already at the three-point level, which were shown \cite{Boulanger:2015ova} to result from the incompleteness of \cite{Vasiliev:1990cm}. The latter is in full agreement with the basic non-existence of such theories \cite{Bekaert:2015tva,Maldacena:2015iua,Sleight:2017pcz,Ponomarev:2017qab}. It is worth mentioning an interesting approach developed in \cite{deMelloKoch:2018ivk,Aharony:2020omh}, which, to some extent, allows one to rewrite the CFT partition function as a path integral with a (very nonlocal) quasi-action for some fields in the bulk. It would be interesting to extend this construction to Chern--Simons vector models. In the same vein \cite{Bekaert:2015tva} one can try to reconstruct the dual theory by ``inverting" the correlation functions. However, such vertices are too nonlocal for the standard field theory tools to work \cite{Bekaert:2015tva,Sleight:2017pcz,Ponomarev:2017qab}. } It seems that AdS/CFT uncovers the same problems as with HiSGRA in flat space, see e.g. \cite{Bekaert:2010hp,Dempster:2012vw,Roiban:2017iqg,Ponomarev:2017nrr}, but they can be identified much quicker thanks to AdS/CFT duality. In this vein HiSGRA are closer to string theory than to field theories \cite{Maldacena:2015iua}.

In the view of the notorious nonlocality problem described above, Chiral Theory seems to be the only local class of HiSGRA with propagating massless fields in four dimensions.\footnote{Any model of quantum gravity should be nonlocal in some sense. Therefore, the nonlocality of HiSGRA is not unexpected. What is lacking at the moment are concrete rules to deal with it. It is clear that the standard field theory tools are inapplicable. Therefore, it is encouraging that there exist some local HiSGRA that can be treated as useful approximations or starting points. Locality comes at a price, of course. } In this paper we construct a manifestly covariant form of equations of motion for Chiral Theory with cosmological constant. The equations are found in the form of a Free Differential Algebra by a smooth deformation of those proposed recently in \cite{Skvortsov:2022syz,Sharapov:2022faa}. This FDA can also be interpreted as  the dual of  an $L_\infty$-algebra. Chirality  makes the theory local both in flat and (anti)-de Sitter spaces, which follows immediately from  our construction. 

Since the bulk of the paper is mostly devoted to the technical aspects of the FDA, it makes sense to discuss conceptual consequences of the results for Chern--Simons Matter Theories and three-dimensional bosonization duality \cite{Giombi:2011kc, Maldacena:2012sf, Aharony:2012nh,Aharony:2015mjs,Karch:2016sxi,Seiberg:2016gmd} right now. Let us recall that the \begin{wrapfigure}{r}{0.37\textwidth}
    \includegraphics[trim={1cm 0.9cm 1cm 0},clip,width=0.37\textwidth]{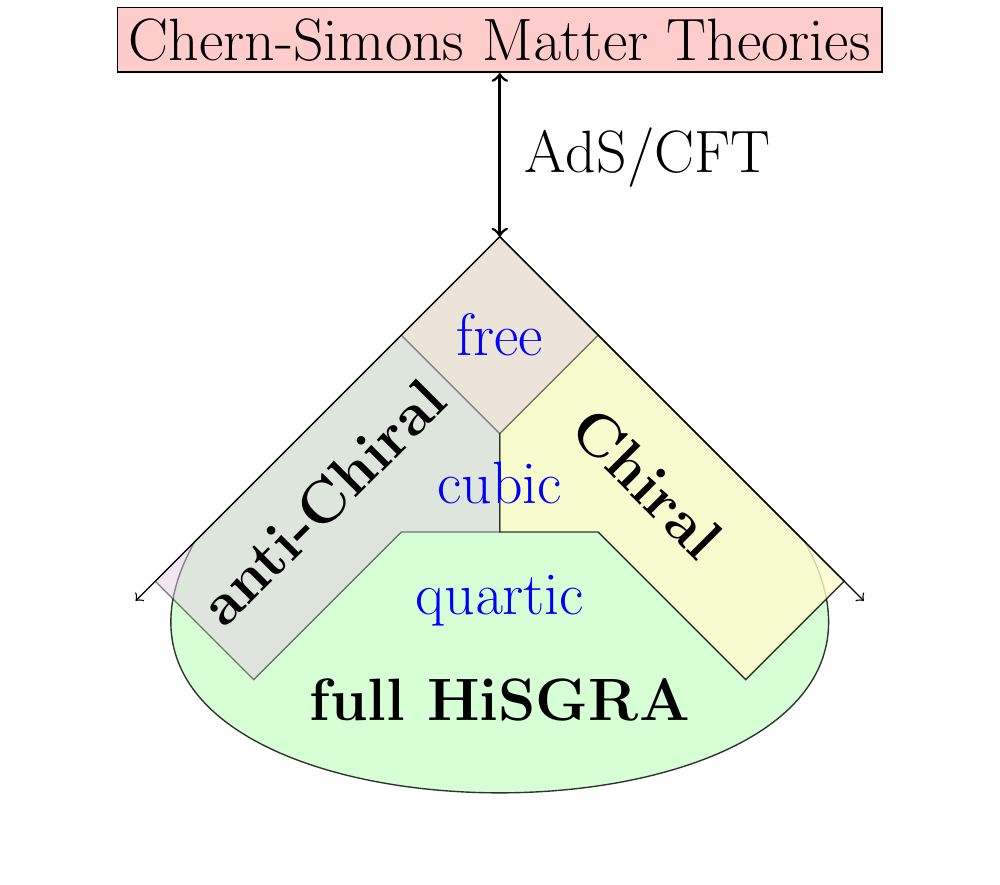}
\end{wrapfigure}
spectrum of Chiral Theory, massless fields with spins $s=0,1,2,3,...$ or $s=0,2,4,...$,\footnote{Gauging Yang--Mills groups is also possible \cite{Skvortsov:2020wtf}. They, however, come in a very restricted Chan-Paton pattern, which is implied by the AdS/CFT duality with vector models. } is the right one to be the dual of Chern--Simons vector models. At the free level, (anti)-Chiral theories and the dual of Chern--Simons vector models coincide. Chiral Theory is smaller than this hypothetical dual HiSGRA in that only the chiral interactions are kept (those where the sum of the helicities on external lines is, say, positive).\footnote{Chiral Theory shares many features with self-dual theories. Some key properties of any self-dual theory include: all solutions of self-dual theories are solutions of the full ones; the same is true for amplitudes. Therefore, self-dual theories are very useful for understanding the complete ones. What is missing at present for the Chiral Theory case is an explicit construction of the complete one.} 

The first consequence of the very existence of a complete, local, Lorentz-covariant theory in $AdS_4$ is that Chern--Simons vector models must have a closed subsector as well. Indeed, one can recover this subsector by computing the holographic correlation functions in Chiral Theory. It would be very interesting to identify this closed subsector directly on the CFT side, which is perhaps hard to achieve at the action level. 

Secondly, the chiral and anti-chiral interactions cover all possible cubic vertices. Now, it is very important that the (anti-)Chiral Theory is rigid and can be thought of as a rock hard building block (all couplings are fixed by higher spin symmetry). Therefore, if one wishes to get all possible three-point functions in the complete unitary HiSGRA dual of Chern--Simons matter theories one just needs to study how to glue chiral and anti-chiral parts together. It turns out that this gluing depends on one additional parameter \cite{Skvortsov:2018uru}, which can be introduced via $U(1)$ electromagnetic duality transformation in the bulk. This immediately proves the three-dimensional bosonization duality at this order since the bulk considerations are insensitive to whether the dual operators are built out of bosonic or fermionic matter. This also gives explicitly the three-point functions \cite{Skvortsov:2018uru}, which are consistent with \cite{Maldacena:2012sf} and explain in a simple way the general structure observed in \cite{Maldacena:2012sf}. 

Thirdly, despite the lack of control over nonlocalities in the full HiSGRA dual to Chern--Simons vector models, existence of its Chiral truncations supports the bosonization duality beyond three-point level. Indeed, the main idea of \cite{Skvortsov:2018uru} is to decompose correlation functions (equivalently, bulk interactions) into parts with definite helicity structure. After the surgery, these building blocks can be rotated with a $U(1)$-phase and sewed back into unitary, but parity violating, correlation functions. The very existence of one-parameter families of conformal field theories that interpolate between parity-preserving bosonic/fermionic vector models can be attributed to the existence of Chiral Theory. It allows one to interpret correlation functions of gauge invariant operators in Chern--Simons vector models as perturbations of the chiral subsector, which effectively introduces one more parameter on top of $1/N$. This is very similar to how Yang-Mills theory can be recast as perturbation of SDYM, likewise for gravity vs. SDGR, the main difference being in that Chiral Theory has its own coupling constant. Eventually these ideas should merge with the pure CFT considerations based on the concept of the slightly-broken higher spin symmetry \cite{Maldacena:2012sf}, where it is possible to prove that there exists a unique class of invariants of this symmetry to serve as correlation functions \cite{Sharapov:2018kjz,Gerasimenko:2021sxj,Sharapov:2022eiy}. 

Lastly, there are good reasons to expect Chiral Theory to be integrable \cite{Ponomarev:2017nrr}, finite and even one-loop exact \cite{Skvortsov:2020gpn}, and to admit a simple twistor formulation, see \cite{Tran:2021ukl} for the first steps in this direction. Therefore, it should be possible to obtain all-loop results in Chiral Theory and, as a consequence, in the corresponding subsectors of Chern--Simons vector models. Note that the bulk coupling is of order $1/N$. Since the results are exact, they should be valid for small values of $N$, which are of phenomenological significance. Hopefully, Chiral Theory can give an example of a bulk theory that is completely solvable.

While it would be very interesting to explore the ideas outlined above in the future, we proceed now to bitter technicalities of constructing the FDA of Chiral Theory, which is necessary to support them. After a brief overview of the formalism in Section \ref{sec:initialdata} we proceed to the main results in Section \ref{sec:FDA}, which are supported by a number of technical Appendices. The main conclusions of the paper have already been given above and a more technical summary and discussion of the results can be found in Section \ref{sec:disco}.

\section{Initial data}
\label{sec:initialdata}
As it was discussed, in the light-cone gauge there is a complete action for Chiral Theory that terminates at cubic interactions in flat space. Different parts (contractions in the language of \cite{Ponomarev:2017nrr}) of the covariant action for Chiral Theory, both in flat and $(A)dS_4$, were constructed in \cite{Krasnov:2021nsq}. The covariant version of Chiral Theory can be defined as a unique Lorentz invariant and local completion of the free action\footnote{As always, we use the same letter to denote a group of symmetric or to be symmetrized indices, e.g. $A_1...A_k\equiv A(k)$. The indices are raised and lowered with the help of $\epsilon^{AB}=-\epsilon^{BA}$ as in the classical book \cite{penroserindler}. }
\begin{align}\label{niceaction}
    S= \int \Psi^{A(2s)}\wedge H_{AA}\wedge \nabla \omega_{A(2s-2)}\,,
\end{align}
where $H^{AB}\equiv e\fud{A}{C'}\wedge e^{BC'}$ is the basis of self-dual two-forms built out of vierbein $e^{AA'}\equiv e^{AA'}_\mu\, dx^\mu$. The completion is unique provided a genuine higher spin interaction is turned on. For simple Yang--Mills and gravitational interactions of higher spin fields the completion is not unique, but admits very simple actions \cite{Krasnov:2021nsq}. The dynamical fields are zero-form $\Psi^{A(2s)}$ and one-form $\omega^{A(2s-2)}\equiv \omega^{A(2s-2)}_\mu\, dx^\mu$. The action can be shown to describe a massless spin-$s$ particle, where the (conventionally) positive helicity, $+s$, mode belongs to $\omega^{A(2s-2)}$ and the negative heliticy, $-s$, resides in $\Psi^{A(2s)}$. On flat, $(A)dS$ and, more generally, any self-dual background the action enjoys the gauge symmetry 
\begin{align}\label{lin-gauge}
    \delta \omega^{A(2s-2)}&= \nabla \xi^{A(2s-2)} +e\fud{A}{C'} \eta^{A(2s-3),C'}\,,& \delta\Psi^{A(2s)}&=0\,,
\end{align}
where $\xi^{A(2s-2)}$ and $\eta^{A(2s-3),C'}$ are zero-forms. 

We refer the reader to \cite{Skvortsov:2022syz,Sharapov:2022faa} for a detailed discussion of the formalism as well as to the original papers \cite{Vasiliev:1986td,Vasiliev:1988sa} which correctly identify the free data and bits of interactions. We will be looking for covariant equations of motion in the form of a Free Differential Algebra. This implies the well-known extension \cite{Vasiliev:1986td,Vasiliev:1988sa} of the dynamical fields with some auxiliary ones. It is convenient to package all the auxiliary and dynamical fields into a pair of master fields: the one-form 
\begin{align*}
    \omega(y,\bry)&= \sum_{n+m=\text{even}}\tfrac{1}{n!m!} \omega_{A(n),A'(m)}\, y^A...y^A\, \bry^{A'}...\bry^{A'} 
 \end{align*}  
 and the zero-form
 \begin{align*}
 C(y,\bry)&= \sum_{n+m=\text{even}}\tfrac{1}{n!m!} C_{A(n),A'(m)}\, y^A...y^A\, \bry^{A'}...\bry^{A'}\,.
\end{align*}
The restriction that the sum $n+m$ be even is the bosonic projection. The (generating functions of) dynamical fields of \eqref{niceaction} are identified with $\Psi(y)= C(y,\bry=0)$ and $\omega(y)=\omega(y,\bry=0)$. The zero-form $C$ also contains the scalar field $C(0,0)$, which is necessarily present in Chiral Theory. With this field content the most general FDA reads
\besubeqs\label{eq:chiraltheory}
\begin{align} 
    d\omega&= \mathcal{V}(\omega, \omega) +\mathcal{V}(\omega,\omega,C)+\mathcal{V}(\omega,\omega,C,C)+...\,,\\
    dC&= \mathcal{U}(\omega,C)+ \mathcal{U}(\omega,C,C)+... \,.
\end{align}
\esubeqs
The formal consistency of the equations, resulting from the nilpotency of the exterior differential $d$, implies that the multilinear maps on the r.h.s. satisfy the $L_\infty$-relations. 
These equations must have $(A)dS_4$ as an exact solution with $C=0$ and 
\begin{align}\label{adsfour}
    \omega_0&= \tfrac14 \omega^{AB}\, y_A y_B+ \tfrac12 e^{AA'}\, y_A\bry_{A'}+\tfrac14 \omega^{A'B'}\, \bry_{A'} \bry_{B'}\,,
\end{align}
where $\omega$'s are (anti-)self-dual parts of the spin-connection. Expanding \eqref{eq:chiraltheory} over $\omega_0$ and picking up the linear terms, one gets the free equations 
\besubeqs\label{linearizeddataA}
\begin{align}
    d\omega &= \mathcal{V}(\omega_0, \omega) + \mathcal{V}(\omega, \omega_0)+\mathcal{V}(\omega_0,\omega_0,C)\,,& 
    d C&= \mathcal{U}(\omega_0, C)\,.
\end{align}
\esubeqs
The parts of $\mathcal{V}(\omega_0, \omega)$, $\mathcal{V}(\omega, \omega_0)$ and $\mathcal{U}(\omega_0, C)$ that contain the spin-connection must act canonically, i.e. they should transform the components of $\omega$ and $C$ as spin-tensors. Therefore, they can be moved to the l.h.s. and combined with $d$ into the Lorentz covariant derivative\footnote{It should be noted  that in the flat limit $\omega^{AB}$ does not play the role of the spin-connection anymore, but $\omega^{A'B'}$ does.} $\nabla$. The free equations must read
\begin{align}\label{linearizeddata}
    \nabla\omega &= e^{BB'}(\hhbar\, \bry_{B'} \pl_{B}+y_{B} \plb_{B'}) \omega +H^{B'B'} \plb_{B'}\plb_{B'}C(y=0,\bry)\,,\\
    \nabla C&= e^{BB'}(\hhbar\, y_{B} \bry_{B'}-\pl_B \plb_{B'}) C\,.
\end{align}
Here we introduced the parameter $\hhbar$ associated with the cosmological constant.  At $\hhbar=0$ we recover the free equations of Chiral Theory in flat space \cite{Skvortsov:2022syz,Sharapov:2022faa}.

\section{FDA}
\label{sec:FDA}
We first summarize the known results about the leading order, then discuss the next-to-leading order vertices,  which are still easy to find by hand, and finally  explain how to systematically generate all higher vertices.  

\subsection{LO}
\label{sec:lowest}

\paragraph{Higher spin algebra.} The $L_\infty$-relations, or the formal consistency of \eqref{eq:chiraltheory}, imply that $\mathcal{V}(f, g)$ satisfies the Jacobi identity and, thereby, $\omega$ takes values in some Lie algebra. As it was understood in different contexts by many authors \cite{Dirac:1963ta, Gunaydin:1981yq,Gunaydin:1983yj,Fradkin:1986ka,Vasiliev:1986qx, Gunaydin:1989um} this Lie algebra originates, in fact, from an associative algebra $A_2^e$, which is the even subalgebra of the second Weyl algebra. In other words, we can start with the four  generators $\hat Y^\aA=(\hat y^A, {\hat \bry}{}^{A'})$ subject to the commutation  relations $[\hat Y^\aA,\hat Y^\aB]=2C^{\aA\aB}$,  with $C^{\aA\aB}$ being the matrix of the canonical symplectic form. The second Weyl algebra $A_2$ is then the associative algebra of polynomials $f(\hat Y)$ in $\hat Y$'s. A useful realization of $A_2$ is via deformation quantization: one considers polynomials $f(Y)$ in the commutative variables $Y^\aA$ that are equipped with the star-product
\begin{align}
    (f\star g)(Y)&= \exp[ Y\cdot \pl^1 +Y\cdot \pl^2 +C^{\aA\aB} \pl^1_\aA \pl^2_\aB]\, f(Y_1)g(Y_2)\Big|_{Y_{1,2}=0}\,,
\end{align}
where $Q\cdot P\equiv Q^\aA P_{\aA}$.
The higher spin algebra is then the even subalgebra $A_2^e$ of $A_2$. It is also convenient \cite{Vasiliev:1988sa} to tensor it with the matrix algebra $\mathrm{Mat}_N$ and define $\hs= A_2^e \otimes \mathrm{Mat}_N$. The trick with the matrix extension is well-justified by AdS/CFT correspondence and by the fact that one can introduce Yang--Mills-type interactions into Chiral Theory \cite{Skvortsov:2020wtf}. 

At the technical level, the matrix extension implies that the $L_\infty$-algebra we are looking for results from an $A_\infty$-algebra via the symmetrization map. For example, the Lie bracket is the commutator in associative algebra $\hs$. As we will see, for technical reason, it is better to define $\hs=A_1 \otimes \mathcal{B}$, where $A_1$ is the (star-product) Weyl algebra in $y^A$ and $\mathcal{B}$ is another associative algebra, which is $A_1\otimes \mathrm{Mat}_N $ in our case (supplemented at the end with the projection onto the even subalgebra of $\hs$). Therefore, we will treat $\omega$ and $C$ as function of $y$'s that take values in some associative algebra and the order of the arguments in vertices is important. With this convention in mind the product in the higher spin algebra can be written as
\begin{align}\label{hsalgebra}
    \mathcal{V}(f,g)&= \exp{[p_{01}+p_{02}+\hhbar\, p_{12}]}f({y}_1)\, g({y}_2)\Big|_{{y}_i=0} \equiv (f\star g)(y)\,,
\end{align}
where we introduced $y^A\equiv p_0^A$, $\pl^{y_i}_{A}\equiv p_{A}^i$. Contractions $p_{ij}\equiv p_i \cdot p_j\equiv -\epsilon_{AB}p^A_{i}p_{j}^B=p^A_{i}p_{jA}$ are such that $\exp[p_0\cdot p_i]f(y_i)=f(y_i+y)$ is the translation operator. More generally, vertices can be represented as poly-differential operators
\begin{align}
    \mathcal{V}(f_1,...,f_n)&= \mathcal{V}(y, \pl_1,...,\pl_n)\, f_1(y_1)...f_n(y_n) \Big|_{y_i=0}\,,
\end{align}
where the $f$'s stand either for $\omega$'s or $C$'s. Occasionally, we will use $q$'s for poly-differential operators in $\bry$, e.g. $\bry^{A'}\equiv q_0^{A'}$, $\pl^{\bry_i}_{A'}\equiv q_{A'}^i$. Most of the time, we omit $|_{y_i=0}$ as well as the arguments of the vertices and write down only the corresponding symbols. 

Coming back to \eqref{hsalgebra}, we introduced $\hhbar$ in order to have a smooth deformation that starts from Chiral Theory with vanishing cosmological constant. Basically, $\hhbar$ is the cosmological constant. In the flat limit ($\hhbar\rightarrow 0$) the first Weyl algebra $A_1$ turns back into the commutative algebra of polynomial functions $\mathbb{C}[y]$, whose quantization $A_1$ is. Now, we can check that the boundary conditions imposed by the free equations \eqref{linearizeddata} are indeed satisfied:
\besubeqs\label{bcomega}
\begin{align}
        \mathcal{V}(e,f)-\mathcal{V}(f,e)&=2 e^{p_{02}+q_{02}} \left(\hhbar\,  q_{01} p_{12}+p_{01} q_{12}\right) (e^{CC'}y^1_C \bry^1_{C'}) f(y^2,\bry^2) \Big|_{y_i,\bry_i=0}\,,\\
        \mathcal{V}(\omega,f)-\mathcal{V}(f,\omega)&=2 \hhbar\,  p_{01} p_{12}\, e^{p_{02}+q_{02}} (\omega^{AB}y^1_A y^1_{B}) f(y^2,\bry^2) \Big|_{y_i,\bry_i=0}\,,\\
        \mathcal{V}(\bar\omega,f)-\mathcal{V}(f,\bar\omega)&=2 q_{01} q_{12}\, e^{p_{02}+q_{02}} (\omega^{A'B'}\bry^1_{A'} \bry^1_{B'}) f(y^2,\bry^2) \Big|_{y_i,\bry_i=0}\,,
\end{align}
\esubeqs
where $f$ is a zero-form. Here an unimportant factor of $2$ appears as a result of our normalization of the higher spin algebra product. The minus in between the terms is due to the fact that argument $f$ corresponds to the one-form $\omega$ which anti-commutes to $e$. It is also clear that \eqref{adsfour} is an exact solution that defines $(A)dS_4$ since $L_{AB}=-\tfrac12 y_A y_B$, $L_{A'B'}=-\tfrac12 \bry_{A'} \bry_{B'}$ and $P_{AA'}=-\tfrac12 y_A \bry_{A'}$ form $(A)dS_4$ algebra with the most important relation being
\begin{align}\label{adsdeform}
    [P_{AA'},P_{BB'}]_\star &= L_{AB} \epsilon_{A'B'} + \hhbar \epsilon_{AB}L_{A'B'}\,.
\end{align}

\paragraph{Dual module. } With the help of the matrix trick the next $A_\infty$-algebra relation implies that $\mathcal{U}(a,u)$ and $\mathcal{U}(u,a)$ define a bimodule $M$ over $\hs$, $a\in \hs$, $u\in M$. It is easy to see that $M$ should be dual to the algebra $\hs$, viewed as a bimodule over itself,  with respect to the following non-degenerate paring:
\begin{align}
    \langle a|u \rangle&= \exp[p_{12}]\,a(y_1)\,u(y_2) \big|_{y_i=0}\,.
\end{align}
The trace over the matrix and, possibly, other factors is understood. The induced bimodule structure is given by the relations
\begin{equation}
    \begin{split}
        &\mathcal{U}_1(\omega,C)=+\exp{[\hhbar\, p_{01}+ p_{02}+p_{12}]}\, \omega({y}_1)\, C({y}_2)\Big|_{\bar{y}_i=0}\,,\\
        &\mathcal{U}_2(C,\omega)=-\exp{[p_{01}-\hhbar\, p_{02}-p_{12}]}\, C({y}_1)\, \omega({y}_2)\Big|_{{y}_i=0}\,.
    \end{split}
\end{equation}
At this point one might want to introduce a reflection operator $R$, $R^2=1$, $R y^A R=-y^A$, so that we can redefine $\mathcal{U}_2$ to have only $+$ in the second exponent at the price of having $C R$ everywhere instead of $C$ (when $R$ is dragged through $\omega$ it effectively flips the sign of all $(\_ \cdot p_2)$-terms. For $\hhbar=1$ the bimodule structure above coincides with the twisted-adjoint action \cite{Vasiliev:1999ba}, but the latter does not have the flat limit. At $\hhbar\neq1$ the left/right actions are different, but are coordinated to give a bimodule. In this sense the right action is derivable from the left one. 

We can again check that the boundary conditions provided by \eqref{linearizeddata} are satisfied:
\besubeqs
\begin{align}
        \mathcal{U}(e,f)+\mathcal{U}(f,e)&=2 e^{p_{02}+q_{02}} \left(\hhbar\,  p_{01} q_{01}+p_{12} q_{12}\right) (e^{CC'}y^1_C \bry^1_{C'}) f(y^2,\bry^2) \Big|_{y_i,\bry_i=0}\,,\\
        \mathcal{U}(\omega,f)+\mathcal{U}(f,\omega)&=2 \hhbar\,  p_{01} p_{12}\, e^{p_{02}+q_{02}} (\omega^{AB}y^1_A y^1_{B}) f(y^2,\bry^2) \Big|_{y_i,\bry_i=0}\,,\\
        \mathcal{U}(\bar\omega,f)+\mathcal{U}(f,\bar\omega)&=2 q_{01} q_{12}\, e^{p_{02}+q_{02}} (\omega^{A'B'}\bry^1_{A'} \bry^1_{B'}) f(y^2,\bry^2) \Big|_{y_i,\bry_i=0}\,,
\end{align}
\esubeqs
with $f$ being a zero-form. The action of the Lorentz subalgebra, i.e. the terms with $\omega^{AB}$ and $\omega^{A'B'}$ are exactly the same as in \eqref{bcomega}. 

\subsection{NLO}
\label{sec:lowest}
The cubic vertices satisfy simple linear equations and can be solved by brute-force, see e.g. \cite{Vasiliev:1988sa,Sharapov:2017yde,Skvortsov:2022syz}, which we closely follow below.  

\paragraph{Cubic Vertex $\boldsymbol{\mathcal{V}(\omega,\omega,C)}$.} In accordance with different orders of the arguments there are three $A_\infty$-maps to be found:
\begin{align}
    \mathcal{V}(\omega,\omega,C)=\mathcal{V}_1(\omega,\omega,C)+\mathcal{V}_2(\omega,C,\omega)+\mathcal{V}_3(C,\omega,\omega)    \,.
\end{align}
The defining relations of an $A_\infty$-algebra requre then  
\begin{equation}
    \begin{split}
        &\mathcal{V}_1(\mathcal{V}(\omega,\omega),\omega,C)-\mathcal{V}(\omega,\mathcal{V}_1(\omega,\omega,C))+\mathcal{V}_1(\omega,\omega,\mathcal{U}_1(\omega,C))-\mathcal{V}_1(\omega,\mathcal{V}(\omega,\omega),C)=0\,,\\
        &\mathcal{V}(\mathcal{V}_1(\omega,\omega,C),\omega)+\mathcal{V}_1(\omega,\omega,\mathcal{U}_2(C,\omega))+\mathcal{V}_2(\mathcal{V}(\omega,\omega),C,\omega)-\mathcal{V}(\omega,\mathcal{V}_2(\omega,C,\omega))\\
        &-\mathcal{V}_2(\omega,\mathcal{U}_1(\omega,C),\omega)=0\,,\\
        &\mathcal{V}(\mathcal{V}_2(\omega,C,\omega),\omega)-\mathcal{V}_2(\omega,C,\mathcal{V}(\omega,\omega))-\mathcal{V}_2(\omega,\mathcal{U}_2(C,\omega),\omega)-\mathcal{V}(\omega,\mathcal{V}_3(C,\omega,\omega))\\
        &+\mathcal{V}_3(\mathcal{U}_1(\omega,C),\omega,\omega)=0\,,\\
        &\mathcal{V}(\mathcal{V}_3(C,\omega,\omega),\omega)-\mathcal{V}_3(C,\omega,\mathcal{V}(\omega,\omega))+\mathcal{V}_3(C,\mathcal{V}(\omega,\omega),\omega)+\mathcal{V}_3(\mathcal{U}_2(C,\omega),\omega,\omega)=0\,.
    \end{split}
\end{equation}
Here one should replace the three $\omega$'s with three different elements of $\hs$ while keeping the order the same in all terms.
One can rewrite these relations in terms of the corresponding symbols, which reduces them  to simple algebraic equations. For example, the first relation gives
\begin{align*}
0&=-\mathcal{V}_1(p_0+\hhbar\, p_1,p_2,p_3,p_4)e^{p_{01}}+\mathcal{V}_1(p_0,p_1+p_2,p_3,p_4)e^{\hhbar\, p_{12}}\\&\qquad -\mathcal{V}_1(p_0,p_1,p_2+p_3,p_4)e^{\hhbar\, p_{23}}+\mathcal{V}_1(p_0,p_1,p_2,\hhbar\, p_3+p_4)e^{p_{34}}\,.
\end{align*}
A nontrivial solution to these equations that is consistent with the $\hhbar\rightarrow0$ limit of \cite{Sharapov:2022faa} reads
\begin{align*}
     \mathcal{V}_1(\omega,\omega,C)&=+p_{12}\, \int_{\Delta_2}\exp[\left(1-t_1\right) p_{01}+\left(1-t_2\right) p_{02}+t_1 p_{13}+t_2 p_{23} +\hhbar (1+t_1-t_2) p_{12} ]\,, \\
     \mathcal{V}_2(\omega,C,\omega)&=-p_{13}\, \int_{\Delta_2}\exp[\left(1-t_2\right) p_{01}+\left(1-t_1\right) p_{03}+t_2 p_{12}-t_1 p_{23}+\hhbar (1-t_1-t_2) p_{13}]\\
       &\phantom{=}\,-p_{13}\, \int_{\Delta_2}\exp[\left(1-t_1\right) p_{01}+\left(1-t_2\right) p_{03}+t_1 p_{12}-t_2 p_{23}+\hhbar (1-t_1-t_2) p_{13}]\,,
    \\
    \mathcal{V}_3(C,\omega,\omega)&=+p_{23}\, \int_{\Delta_2}\exp[\left(1-t_2\right) p_{02}+\left(1-t_1\right) p_{03}-t_2 p_{12}-t_1 p_{13}+\hhbar (1+t_1-t_2) p_{23} ]\,.
\end{align*}
Hereinafter we let $\Delta_n$ denote the $n$-dimensional simplex $t_0=0\leq t_1\leq ...\leq t_n\leq 1$. This vertices deform smoothly those of Chiral Theory in flat space \cite{Sharapov:2022faa} and are simple $\hhbar$-modifications of what can be extracted from \cite{Vasiliev:1999ba}. 

It is significant that all $\mathcal{U}$-vertices can be obtained from $\mathcal{V}$-vertices by the following duality relation: 
\begin{equation}\label{DR}
    \langle \mathcal{V}(\omega,\omega,C,\ldots,C)|C\rangle=\langle \omega|\mathcal{U}(\omega, C,\ldots,C)\rangle\,.
\end{equation}
 It can be shown that this recipe gives consistent $\mathcal{U}$-vertices provided one carefully keeps track of the ordering. For example, at the cubic order we still need three $\mathcal{U}$-vertices
\begin{align}
    \mathcal{U}(\omega,C,C)=\mathcal{U}_1(\omega,C,C)+\mathcal{U}_2(C,\omega,C)+\mathcal{U}_3(C,C,\omega)   \,.
\end{align}
By duality we find
\begin{align}
    \mathcal{U}_1(p_0,p_1,p_2,p_3)&=+ \mathcal{V}_1(-p_3,p_0,p_1,p_2)\,,\\
    \mathcal{U}_2(p_0,p_1,p_2,p_3)&=- \mathcal{V}_2(-p_1,p_2,p_3,p_0)\,,\\
    \mathcal{U}_3(p_0,p_1,p_2,p_3)&= -\mathcal{V}_3(-p_1,p_2,p_3,p_0)\,.
\end{align}
It is interesting to have a look at least at one of them:
\begin{align}
    \mathcal{U}_1&=p_{01} \int_{\Delta_2}\exp \left[\hhbar\left(t_1-t_2+1\right)   p_{01}+t_1 p_{02}+\left(1-t_1\right) p_{03}+t_2 p_{12}+\left(1-t_2\right) p_{13}\right]\,.
\end{align}
The most important observation is that this vertex is local, as it will be explained in more detail below.\footnote{As a historical comment, let us point out that at $\hhbar=1$ there is a well-known twisted-adjoint interpretation \cite{Vasiliev:1999ba} of the bilinear $\mathcal{U}$-maps. From this viewpoint, it is possible to generate all $\mathcal{U}$-vertices from the $\mathcal{V}$-vertices by the schematic formula  $\mathcal{U}=CR \delta\mathcal{V}/\delta \omega$, $R$ being the reflection operator. This, however, leads to nonlocal vertices immediately. } This is thanks to it not having $p_{23}$ inside the exponent.

\subsection{\texorpdfstring{N${}^k$LO}{Higher orders}}
\label{sec:higher}
\paragraph{Set up.} Following the original ideas of \cite{Vasiliev:1990cm,Vasiliev:1999ba}, but with important modifications of \cite{Sharapov:2022faa} that make the construction to yield local interactions,\footnote{In the $L_\infty$-language the idea of \cite{Vasiliev:1990cm,Vasiliev:1999ba} is to construct an anti-minimal model, i.e. a differential graded Lie algebra, whose minimal model gives the sought-for $L_\infty$. This approach faces the usual problems of (non)-locality \cite{Boulanger:2015ova,Skvortsov:2015lja}, see also below. Therefore, it is remarkable that a simple modification proposed in \cite{Sharapov:2022faa} gives well-defined local vertices of Chiral Theory in flat space and it is possible to turn on the cosmological constant as we show in the paper. The modifications include: a star-product, dual module structure for $C$, the duality map to get $\mathcal{U}$-vertices. } we define an extended algebra of polynomial functions $\mathbb{C}[y,z]$, $Y^a\equiv (y^A, z^A)$, that is equipped with the star-product defined by the matrix  
\begin{align}\label{PB}
   (\Omega^{ab})= -\begin{pmatrix}
        \hhbar \epsilon & \epsilon \\
        -\epsilon & 0
    \end{pmatrix}\,,\qquad \epsilon\equiv \epsilon^{AB}\,.
\end{align}
The symbol of $\mu(f,g)\equiv(f\star g)(y,z)$ is 
\begin{align}
    \exp{[ p_{01}+p_{02}+r_{01}+r_{02} +p_1\cdot r_2-r_1\cdot p_2 +\hhbar\, p_{12}]}\,.
\end{align}
Hereinafter $r$ for $z$ is the same as $p$ for $y$.
The integral representation reads\footnote{As in \cite{Sharapov:2022faa} we define the integral in such a way that $\int d^2u\, \exp[ u\cdot q]=\delta^2(q)$.}
\begin{align}
    (f\star g)(y,z)&= \int du\,dv\,dp\,dq\,f(y+u,z+v) g(y+q,z+p) \exp{[v\cdot q-u\cdot p +\hhbar\, p\cdot v]}\,.
\end{align}
The vertices are generated via the standard Homological Perturbation Theory. We refer to \cite{Sharapov:2022faa} for a detailed description and to Appendix \ref{app:algebra} for the necessary modifications. Below, we just explain the algorithm  and illustrate it with some examples. The generators of the algebra $y_A$, $z_A$ act as follows:
\begin{align}\label{starrrr}
   & \begin{aligned}
    y_{A}\star f &=(y_{A}  -\hhbar\, \pl^y_{A}-\pl^z_{A})f\,, \\
    f \star y_{A} &=(y_{A}  +\hhbar\, \pl^y_{A}-\pl^z_{A})f\,, 
    \end{aligned}   &  z_{A}\star f&=f\star z_A= (z_{A}+\pl^y_{A}) f\,.
\end{align}
With these relations one can see that the function $\varkappa = \exp[z^{C}y_{C}]$ obeys the identities 
$$\{y_{A}, \varkappa\}_\star  = z_{A}\star\varkappa =\varkappa\star z_{A}=0\,.$$ 
We will need a further  extension of the above algebra to the algebra of differential forms in $d z$. The corresponding exterior differential will be denoted by $d_z$. We will use the Poincar\'e lemma in the following form:
\begin{align}\label{homofor}
      f^{(1)}=h[f^{(2)}]&= d z^{A}\, z_{A} \int_0^1 t\, dt\, f^{(2)}(t z)\,, &
      f^{(0)}=h[f^{(1)}]&= z^{A} \int_0^1 dt\, f_{A}^{(1)}(t z)\,,
\end{align}
see e.g. \cite{Didenko:2014dwa}.
The first expression is a particular solution to $d_z f^{(1)}= f^{(2)}$ for a one-form $f^{(1)}\equiv dz^{A} f^{(1)}_{A}(z)$ and a two-form $f^{(2)}\equiv \tfrac12 f^{(2)}(z) \epsilon_{AB}dz^{A} dz^{B}$. The second formula is a particular solution to $d_z f^{(0)}= f^{(1)}$ for a closed one-form $f^{(1)}$ and a zero-form $f^{(0)}\equiv f^{(0)}(z)$. We also define $h[f^{(0)}]=0$ for any zero-form $f^{(0)}$. 

The aforementioned Homological Perturbation Theory generates all trees that have $\omega$ and $C$ as leaves. The trivalent vertices of trees are provided by the $\star$-product $\mu$, while the contracting homotopy $h$ defines the internal branches. Given a zero-form $C(y)$, we denote
\begin{equation}
    \delta C =\tfrac12C(z)e^{z^B y_B}dz_A dz^A\,.
\end{equation}
The  zero-form $C$ enters the trees via the combination $\Lambda[C]= h\delta C$, that is, 
\begin{align}
    \Lambda[C]&= dz^{A}z_{A} \int_0^1 t\,dt\, C(tz)e^{tz^By_B}=dz^{A}z_{A} \int_0^1 t\,dt\, \exp[tz\cdot (y+p_i)] \,C(y_i)\Big|_{y_i=0}\,,
\end{align}
where the last expression is the symbol of the previous one that we use in practice. 
All trees that make sense can in principle contribute. It should be remembered that $h^2=0$ and forms of degree higher than two vanish identically. Each subtree of a given tree must be an admissible tree. As was shown in \cite{Sharapov:2022faa}, and remains to be true here, there are certain classes of trees that vanish thanks to an interplay between specific $\mu$ and $h$ we have defined. It is instructive to give some examples. 

\paragraph{NLO.} There are four nontrivial trees that can be drawn at NLO. For example, $\mathcal{V}(\omega,\omega,C)$ is built from
$$
   \mathcal{V}(\omega,\omega,C)=\omega(y) \star h[ \omega(y) \star \Lambda[C] ]= \begin{tikzcd}[column sep=small,row sep=small]
   & {}& \\
    & \mu\arrow[u]  & \\
    \omega\arrow[ur]  & & \mu\arrow[ul, "h" ']   & \\
    & \omega \arrow[ur]& &\Lambda[C]\arrow[ul]
\end{tikzcd}
$$
Evaluation of this tree is almost identical to that of \cite{Sharapov:2022faa}. It is easy to compute the other three trees, two of which contribute to $\mathcal{V}(\omega,C,\omega)$ and the mirror copy of the one above to $\mathcal{V}(C,\omega,\omega)$.  This gives exactly the analytical expressions of Section \ref{sec:lowest}.

\paragraph{NNLO.} There are many more trees at NNLO, which correspond to quartic vertices from the equations of motion vantage point. Nevertheless, there are only two nontrivial topologies:\footnote{Note that, as different from \cite{Sharapov:2022faa}, the $\star$-product $\mu$ is non-commutative. Therefore, one cannot obtain all the vertices by simple permutations of the branches of a few basic trees.}
$$
   G_1=a \star h[ h[ b \star \Lambda[u] ] \star \Lambda[v]]= \begin{tikzcd}[column sep=small,row sep=small]
   &{}&\\
    & \arrow[u] \mu  & \\
    a\arrow[ur]&&  \arrow[ul,"h"']\mu &  \\
    & \arrow[ur,"h"]\mu &&\arrow[ul]\Lambda[v] \\
    \arrow[ur]b & & \arrow[ul]\Lambda[u] &
\end{tikzcd}
$$
and
$$
   G_2= h[ a \star \Lambda[u] ]\star  h[ b \star \Lambda[v]]= \begin{tikzcd}[column sep=small,row sep=small]
   &&&{}&&&\\
    &&& \arrow[u]\mu  &&& \\
    & \mu\arrow[urr,"h"]& && & \arrow[ull,"h"']\mu &  \\
    a\arrow[ur]&& \arrow[ul]\Lambda[u]  & &  b\arrow[ur]&&\arrow[ul]\Lambda[v]
\end{tikzcd}
$$
Here we keep abstract arguments $a$, $b$, $u$, and $v$ instead of $\omega$ and $C$. The first tree is the only contribution to $\mathcal{V}(\omega,\omega,C,C)$. Explicit analytical expressions for $G_1$ and $G_2$ can be found in Appendix \ref{app:NNLO}. They have the following structure:
\begin{align}
    G_1&= \ast (p_{12})^2\exp{[ \ast p_{01}+ \ast p_{02} +\ast p_{13} +\ast p_{23} +\ast  p_{14} +\ast  p_{24} + \ast \hhbar\, p_{12}]}\,,\\
    G_2&=\ast(p_{13})^2 \exp{[\ast p_{01} +\ast p_{03} +\ast p_{12} +\ast p_{23} +\ast p_{14} +\ast p_{34} +\ast \hhbar\, p_{13}]}\,.
\end{align}
Hereinafter $\ast$ stands for unimportant prefactors that depend on $t$'s. The vertices are smooth deformations of those from \cite{Sharapov:2022faa}. Most importantly, they are still local, as will be explained below in more detail. The locality of $G_1$ implies that it does not depend on $p_{34}$ and the locality of $G_2$ means  its independence of $p_{24}$. Out of curiosity we compute a quintic (N${}^3$LO) vertex in Appendix \ref{app:NNNLO} and its higher order cousins.

\paragraph{Locality.} It is important to engineer the vertices that are local. Locality is a crucial difference between the formal deformation procedure and an actual field theory. Indeed, it is well-known that Noether procedure (constructing interactions order by order in the fields, which is usually applied to gauge theories) always admits an all order solution once the locality is abandoned \cite{Barnich:1993vg}. An equivalent statement is even simpler to make in the light-cone gauge: any function can be taken as a Hamiltonian $H$ (provided simple kinematical constraints are imposed) unless we care about locality of the boost generators $J^{i-}$. Within the FDA approach, where the equations $d\Phi=Q(\Phi)$ are determined by an odd nilpotent vector field $Q$, the r.h.s. of the equations can easily lead to meaningless interactions from the field theory point of view, while $Q$ gives a well-defined $L_\infty$-algebra. The problem is that fields $\Phi(x)$ (as maps from spacetime to the target space supermanifold $\mathcal{N}$ where $Q$ is defined) can be expressed as derivatives of each other by virtue of the equations of motion. Therefore, nonlinearities in $Q(\Phi)$ can form infinite series in derivatives.\footnote{A vague statement is that in any reasonable measure on the space of $Q$ for a given $\mathcal{N}$, most of $Q$ lead to nonsensical equations of motion due to nonlocalities. Lucky exceptions to this situation are topological field theories where $\mathcal{N}$ is finite-dimensional. For field theories, and certainly for our case, $\mathcal{N}$ is infinite dimensional. } 

More specifically, nonlocality can develop in any vertex with two or more $C$-fields since they contain auxiliary fields that are expressed as derivatives of the dynamical ones:
\begin{align}
    C^{A(2s+k),A'(k)}&\sim \nabla^{AA'}...\nabla^{AA'} C^{A(2s)}\,, && k=0,1,2,... \,.
\end{align}
Here $C^{A(2s)}\equiv \Psi^{A(2s)}$ is one of the dynamical fields. The same is true for $C^{A(k),A'(2s+k)}$, where $C^{A'(2s)}$ is $s$ derivatives further removed from $\omega^{A(2s-2)}$. For definiteness let us concentrate on $\mathcal{V}(\omega,\omega,C,...,C)$-type vertices with $n$ fields $C$. Nonlocality is present once one has an infinite series of the form\footnote{Not every infinite series in derivatives is ill-behaved as long as $a_k$ decay to zero fast enough, see e.g. examples in \cite{Boulanger:2015ova,Skvortsov:2015lja}. Nevertheless, for Chiral Theory, which is known to be local, having an infinite series in derivatives must be avoided. }
\begin{align}\notag
  \sum_k a_k\overbrace{\nabla_{BB'}...\nabla_{BB'}}^k C^{A(2s)} \nabla^{BB'}...\nabla^{BB'} C^{ M(2s)}...\sim \sum_k a_k C\fud{A(2s)}{B(k),B'(k)}C^{M(2s)B(k),B'(k)}...\,,
\end{align}
where we singled out two $C$-arguments and ignored the rest ones.
In terms of the symbols of operators, nonlocality manifests itself as an infinite series in $w=q_{ij}p_{ij}$ where $i$, $j$ correspond to any pair of $C$-arguments in $\mathcal{V}(\omega,\omega,C,...,C)$. 

The vertices we construct here (and the same is true for those of Chiral Theory without cosmological constant \cite{Sharapov:2022faa}) have a very special form:
\begin{align}
    \mathcal{V}(\omega,\omega,C,...,C)&= \ast \exp[\ast p_{01}+\ast p_{02}+ \ast \hhbar \, p_{12} +\sum_{2<i\leq n+2} \ast p_{1i}+\sum_{2<i\leq n+2} \ast p_{2i}]\,.
\end{align}
Recall that for every vertex there is an overall star-product factor for another copy of $A_1$, in $\bry$:
\begin{align}
    \exp [ \sum_{0\leq i<j\leq n+2} q_{ij}]\,.
\end{align}
Therefore, locality is equivalent to not having $p_{ij}$ inside $\exp[...]$ with $2< i<j\leq n+2$, which is indeed the case. Applying now the duality relation (\ref{DR}), we  find immediately 
\begin{align*}
   &\mathcal{U}(p_0,p_1,...,p_{n+2}) =\mathcal{V}(-p_{n+2},p_0,p_1,...,p_{n+1}) = \\
  & \ast \exp[ \ast p_{0,n+2}+\ast p_{1,n+2}+\ast \hhbar\, p_{01}+\sum_{1<i\leq n+1} \ast p_{0,i}+\sum_{1<i\leq n+1} \ast p_{1,i}]\,.
\end{align*}
It is a local vertex as well! The argument that proves locality to all orders is exactly the same as in \cite{Sharapov:2022faa}. Therefore, we conclude that all the vertices are local and, thereby, define a theory rather than just an $L_\infty$-algebra. It is worth clarifying that by a {\it local theory} we mean a theory where every vertex has a finite number of derivatives provided the helicities on all external legs are fixed.  

\section{Summary and discussion}
\label{sec:disco}
In this paper, we constructed Chiral Theory with cosmological constant, whose existence was anticipated already in \cite{Ponomarev:2016lrm} and supported further in \cite{Metsaev:2018xip,Skvortsov:2018uru} with the help of the light-cone gauge techniques. Technically, the theory is constructed as classical equations of motion in the form of FDA. The FDA emerges from the homological perturbation theory in a standard way. The theory is a straightforward and smooth deformation of Chiral Theory's FDA in flat space \cite{Skvortsov:2022syz,Sharapov:2022faa}, the deformation parameter $\lambda\sim 1/R$ being  the `inverse radius' of (anti-)de Sitter space. It is still remarkable that the self-evident deformation of \cite{Sharapov:2022faa} leads to a local theory.\footnote{Even if we are given a local theory to begin with, it is possible to perform nonlocal field-redefinitions that change physical observables and render them nonsensical \cite{Skvortsov:2015lja,Boulanger:2015ova}, while still having a well-define $L_\infty$-algebra. A (non)local field-redefinition that admits a series expansion in derivatives is just a change of a basis at the level of the $L_\infty$-algebra. Therefore, it is not guaranteed that switching on the cosmological constant the way we did leads to a local theory. There are higher spin examples \cite{Skvortsov:2015lja,Boulanger:2015ova} where this does not happen. Within the homological perturbation theory one deforms a differential of a strong deformation retract and however simple the deformation is it may be hard to predict the structure of the resulting $L_\infty$-algebra. } Despite the conceptual simplicity of the result, it is of significant importance in view of the duality to Chern--Simons Matter Theories, discussed in Introduction.

In view of footnote \ref{vasilkaput}, which recaps the main problems of holographic HiSGRA's, it is worth summarizing what is already known about interactions of these hypothetical HiSGRA's. A nonlocal quartic vertex was holographically reconstructed in \cite{Bekaert:2015tva}, followed by a complete cubic action \cite{Sleight:2016dba}. Within the FDA approach the local form of $\mathcal{U}(\omega,C,C)$, which corrects \cite{Vasiliev:1988sa}, was found in \cite{Didenko:2018fgx}. The best available result is a holomorphic subsector of $\mathcal{V}(\omega,\omega,C,C)$-type interactions \cite{Didenko:2019xzz} and parts of $\mathcal{V}(\omega,C,C,C)$-type interactions \cite{Didenko:2020bxd,Gelfond:2021two}, which contains some of the vertices found previously in  \cite{Metsaev:2018xip,Skvortsov:2018uru} within the light-cone approach. Note that none of the FDA vertices found so far addresses the genuine nonlocality/non-existence problem \cite{Bekaert:2015tva,Maldacena:2015iua,Sleight:2017pcz,Ponomarev:2017qab}, the simplest one being to reproduce \cite{Bekaert:2015tva} by genuine bulk methods without inverting the holographic correlators. Another problem is that each order requires a separate analysis even for the vertices that must be local.

Since the papers above and our deal with higher spin interactions in $(A)dS_4$ and we used the notation close to those of \cite{Vasiliev:1986td,Vasiliev:1988sa,Vasiliev:1990cm,Vasiliev:1999ba}, it is easy to discuss the main differences that allowed us to make progress and to construct an actual theory. Let us recall that the spectrum of $\omega$ and $C$ is the same, since it is determined by the free limit and all higher spin theories, loosely speaking, contain massless fields of all spins. The challenge is to construct interaction vertices, i.e. $A_\infty/L_\infty$-maps as poly-differential operators, that obey locality. There are several ingredients: (a) $y-z$ star-product and contracting homotopy; (b) dual bimodule structure of $C$; (c) duality map. (a) Our star-product \eqref{starrrr} is different from \cite{Vasiliev:1990cm} and \cite{Didenko:2019xzz}: it is fixed, well-defined and does not require any subtle limits, cf. \cite{Didenko:2019xzz}, it works as it is. An invariant characteristic of every star-product is the rank of the underlying Poisson tensor: it is $4$ for \cite{Vasiliev:1990cm} and \cite{Didenko:2019xzz}, $0$ for Chiral Theory in flat space \cite{Sharapov:2022faa} and $2$ for its $(A)dS_4$-deformation of the present paper (which is exactly what is needed to turn on the cosmological constant \eqref{adsdeform} and it closely related to the infinity twistor). It is interesting that the star-product of \cite{Didenko:2019xzz}, while different from ours, gives a local $\mathcal{V}(\omega,\omega,C,C)$-vertex in the limit where the star-product itself is ill-defined. We use the standard contracting homotopy for the de Rham complex of a linear space. Ingredients (b) and (c) are completely new \cite{Sharapov:2022faa} and play the most important part. (b) Our zero-form $C$ takes values in the bimodule dual to the higher spin algebra, rather than in the twisted-adjoint one \cite{Vasiliev:1999ba}. In other words, $C$ is not an element of the higher spin algebra and, for this reason, it should be treated differently and we use the general homological perturbation theory, which cannot be captured by \cite{Vasiliev:1990cm,Didenko:2019xzz}. In particular, this ensures the smooth flat limit or deformation to $(A)dS_4$, which is a counterexample to the general folklore \cite{Vasiliev:1999ba}. (c) It is of crucial importance that $\mathcal{U}$-vertices are obtained via the duality map from $\mathcal{V}$-vertices, which is an original idea born for flat space Chiral Theory \cite{Sharapov:2022faa}, otherwise they emerge nonlocal as in \cite{Vasiliev:1988sa,Vasiliev:1990cm} and require a separate treatment \cite{Didenko:2020bxd,Gelfond:2021two}. To conclude, while one can play with different (a) more or less successfully, it is (b)+(c) that are crucial for constructing a local theory. An even stronger difference is that our main result is a smooth deformation of Chiral Theory in flat space that is already known and well-defined, the deformation having $\lambda$ as a free parameter.

In general, it would be interesting to see what is a maximal local closed sub-sector of the holographic dual of Chern--Simons Matter Theories. We expect that Chiral Theory covers all of it. In this sense, the relation between our results and particular vertices of \cite{Didenko:2018fgx,Didenko:2019xzz,Didenko:2020bxd,Gelfond:2021two} is not clear beyond the lowest orders.\footnote{Cubic vertices decompose into chiral and anti-chiral parts \cite{Metsaev:2018xip}, and hence, different truncations -- holomorphic/chiral/self-dual -- mean essentially the same at this order. Note that the usual cubic vertices from the Lagrangian vantage point appear in $\mathcal{V}(\omega,\omega,C)$, $\mathcal{U}(\omega,C,C)$ and $\mathcal{V}(\omega,\omega,C,C)$. There is an infinite-parameter ambiguity at higher orders \cite{Vasiliev:1999ba,Sharapov:2020quq}, but this analysis does not take locality into account, which may eliminate some parameters as well as to introduce new ones. } It may well be that there are well-defined, in the sense of being local, holomorphic subsectors of \cite{Vasiliev:1999ba} and the question is whether they are smaller/larger than Chiral Theory. One way or another, Chiral Theory in $(A)dS_4$ is directly constructed as a smooth and local deformation of its flat space cousin \cite{Sharapov:2022faa}.\footnote{If Chiral Theory in $(A)dS_4$ corresponds to the local form (not yet available) of the holomorphic vertices of \cite{Vasiliev:1999ba}, our paper gives a `one-line solution' in all orders that provides an alternative completion to the low order analysis of \cite{Didenko:2018fgx,Didenko:2019xzz,Didenko:2020bxd,Gelfond:2021two}. } 

The vertices we obtained turn out to be much simpler than they come out of the homological perturbation theory we use and it would be very interesting to find all of them explicitly as well as to make this hidden simplicity manifest.  An all order result can be found in Appendix \ref{app:NNNLO}. It would also be important to make contact with the Shoikhet--Tsygan--Kontsevich formality and its possible extension to Poisson Orbifolds, which is visible \cite{Sharapov:2017yde,Sharapov:2022eiy} in HiSGRA applications. In particular, it is a challenge to identify the configuration space of the integrals that define vertices and reduce the proof of the $A_\infty/L_\infty$-relations to Stokes theorem \cite{Kontsevich:1997vb, Shoikhet:2000gw}. 

The explicit construction of covariant HiSGRA with massless propagating fields -- Chiral Theory with or without cosmological constant and its contractions \cite{Krasnov:2021nsq} -- opens up new research directions. Among these are (i) computing (higher) holographic correlation functions; (ii) exploring quantum corrections; (iii) constructing exact solutions via the general techniques developed over the years, see e.g. \cite{Sezgin:2005pv,Didenko:2009td,Aros:2017ror} and \cite{Didenko:2021vdb,Didenko:2021vui} for the careful treatment of locality. One might also expect the existence of a simple twistor action for Chiral Theories, see \cite{Hahnel:2016ihf,Adamo:2016ple,Tran:2021ukl} for the twistor results that deal with a contraction of Chiral Theory \cite{Tran:2021ukl} and with conformal HiSGRA \cite{Hahnel:2016ihf,Adamo:2016ple}.

In the light of our results and AdS/CFT duality, there should also exist Chiral Theories with partially-massless fields that are truncations of the holographic duals of higher derivative vector models \cite{Bekaert:2013zya}. Such theories should still admit a smooth flat limit where a free partially-massless field becomes a reducible non-unitary representation of Poincare group. Closely related theory is the self-dual truncation of conformal HiSGRA, which admits a twistor description \cite{Hahnel:2016ihf,Adamo:2016ple}. The latter construction should have a simple generalization to conformal HiSGRA's that are based on higher order singletons \cite{Bekaert:2013zya}. It would also be interesting to construct these two (conjectured) new classes of HiSGRA explicitly in the FDA form. It is also obvious that the FDA we constructed contains contractions \cite{Ponomarev:2017nrr,Krasnov:2021nsq} of Chiral Theory.

\section*{Acknowledgments}
\label{sec:Aknowledgements}
The work of E.S. was partially supported by the European Research Council (ERC) under the European Union’s Horizon 2020 research and innovation programme (grant agreement No 101002551) and by the Fonds de la Recherche Scientifique --- FNRS under Grant No. F.4544.21. A. Sh. gratefully acknowledges the financial support of the Foundation for the Advancement of Theoretical Physics and Mathematics “BASIS”.

\appendix

\section{Some algebra}
\label{app:algebra}
Throughout this section, we deal with partial algebras of different types. Recall that a {\it partial algebra} is just a nonempty set $X$ equipped with a collection $P$ of partial operations on $X$, see e.g. \cite[Ch.2 ]{Gratzler}, \cite{Ljapin}. We will always assume that $X$ is a complex vector space, so that $P$ includes the addition of vectors and multiplication them by complex numbers. The other operations, which we all assume to be multilinear over $\mathbb{C}$, may be defined only for some elements of $X$. If $p\in P$ is an $n$-ary operation, then the {\it domain} of $p$, is a subset $\mathrm{Dom}(p)\subset X\times \cdots\times X$ in the $n$-th Cartesian power of $X$ such that $p$ is defined for all the elements of $\mathrm{Dom}(p)$. In the case that the domain of each operation of $P$ coincides with the whole Cartesian power of $X$, the pair $(X,P)$ is just an ordinary (or total) algebra. In the rest of this section we will omit the adjective `partial'.

Let $\mathfrak{A}=\mathbb{C}[[y,z]]$ be the space of formal power series in four indeterminates $y^A$ and $z^A$, $A=1,2$. We make it into an algebra over $\mathbb{C}$ for the Weyl--Moyal $\star$-product\footnote{As in the main text, the  index $A$ is raised and lowered with the help of the $\epsilon$-symbol and $y\cdot z\equiv y^Az_A$.} 
\begin{equation}\label{sp}
a\star b=\exp\big[\pl_{y_1}\cdot \pl_{z_2} -\pl_{z_1}\cdot \pl_{y_2} +\hhbar \, \pl_{y_1}\cdot \pl_{y_2}\big] a(y_1,z_1) \,b(y_2,z_2)\,\Big|_{\substack{\displaystyle{}y_{1,2}=y \\\displaystyle{}z_{1,2}=z}}\,\,,
\end{equation}
$\hhbar$ being a complex parameter.
Notice that the $\star$-product is strongly associative in the sense of \cite[Ch.1.5]{Ljapin} and its domain includes the Cartesian square of the subspace $\mathbb{C}[y,z]\subset\mathfrak{A}$, i.e., the space of complex polynomials in $y$'a and $z$'s.  

As usual, one may regard the algebra $\mathfrak{A}$ as a bimodule over itself. 
The complex vector space $\mathfrak{A}$ enjoys a $\mathbb{C}$-linear involution $\tau: \mathfrak{A}\rightarrow \mathfrak{A}$ defined by 
\begin{equation}
   a(y,z)\quad \mapsto \quad a^\tau (y,z)=a(z,y)e^{z\cdot y}\,. 
\end{equation}
Clearly, $\tau^2=1$. Using this involution, we give the space $\mathfrak{A}$  another structure of $\mathfrak{A}$-bimodule,  denoted by $\mathfrak{A}^\tau$. By definition,
\begin{equation}\label{Mod}
    a\circ m\circ b=(a\star m^\tau\star b)^\tau\qquad \forall a,b\in \mathfrak{A}\,,\quad \forall m\in \mathfrak{A}^\tau\,.
\end{equation}
The bimodule axioms are verified immediately. Let us introduce the subspace $\mathfrak{M}=\mathbb{C}[[y]]\subset \mathfrak{A}^\tau$. We claim that $\mathfrak{M}$ is actually an $\mathfrak{A}$-submodule. This is enough to check only for the generators. We find
\begin{equation}
\begin{aligned}
     y^A\circ m=(-\partial_y^A+\hhbar y^A)m(y)\,,\\
     m\circ y^A=(-\partial_y^A-\hhbar y^A)m(y)\,,
\end{aligned}
     \qquad \qquad z^A\circ m=m\circ z^A=0\,.
\end{equation}
Notice that the left and right actions of $\mathfrak{A}$ on $\mathfrak{M}$ are different unless $\hhbar\neq 0$. As usual, one can regard the  $\mathfrak{A}$-bimodule $\mathfrak{M}$ as a graded 
algebra $\mathfrak{A}\oplus \mathfrak{M}$ w.r.t. the product
\begin{equation}\label{prod}
    (a,m)(\tilde a,\tilde m)=(a\tilde a, a\tilde m+\tilde a m)\,.
\end{equation}
The elements of the subspaces $\mathfrak{A}$ and $\mathfrak{M}$ are prescribed the 
degrees $0$ and $1$, respectively. 

A map $\bar \delta:\mathfrak{M}\rightarrow \mathfrak{A}$ will define a differential of the 
algebra $\mathfrak{A}\oplus \mathfrak{M}$ of degree $-1$, making it into a dg-algebra, iff the following relations are satisfied:
\begin{equation}\label{brel}
    \bar\delta(a\circ m\circ b)=a\star \bar\delta m\star b\,,\qquad \bar\delta m\circ \tilde m=m\circ\bar\delta \tilde m\,.
\end{equation}
Let us set $\bar\delta m=m^\tau$. Then the first relation takes the form 
\begin{equation}
     (a\circ m\circ b)^\tau=a\star m^\tau\star b\,,
\end{equation}
which is equivalent to the definition (\ref{Mod}). The second relation boils down to the identity
\begin{equation}
    m^\tau\circ \tilde m=(m^\tau\ast \tilde m^\tau)^\tau= m\circ \tilde m^\tau\,.
\end{equation}
As a next step, we introduce the Grassmann algebra $\Lambda$ on two generators  $dz^A$, 
\begin{equation}
    dz^Adz^B=-dz^B dz^A\,,
\end{equation}
both in degree $1$, and define the tensor product algebra $\mathfrak{B}=\mathfrak{A}\otimes \Lambda$. Assuming the generators $dz^A$ act trivially in $\mathfrak{M}$, i.e.,
\begin{equation}
    dz^A \circ m=m\circ dz^A=0\,,
\end{equation}
we endow $\mathfrak{M}$ with the structure of 
$\mathcal{B}$-bimodule and define the graded algebra $\mathfrak{B}\oplus \mathfrak{M}$ with the product (\ref{prod}), where now $(a,m)\in \mathfrak{B}\oplus\mathfrak{M}$. The differential $\bar \delta$ above gives rise to a differential $\delta$ of degree $+1$ on 
$\mathfrak{B}\oplus \mathfrak{M}$. The latter is defined as
\begin{equation}
   \delta a=0 \,,\qquad \delta m = m^\tau dz^1  dz^2\in \mathfrak{B}
\end{equation}
for all $a\in \mathfrak{B}$ and $m\in \mathfrak{M}$. It easy to see that relations (\ref{brel}) hold true for $\mathfrak{A}$ replaced with $\mathfrak{B}$ and $\bar \delta$ replaced with $\delta$. Hence, we may regard $\mathfrak{B}\oplus \mathfrak{M}$ as a dg-algebra with differential $\delta$. 

Considering now the elements $a(y,z,dz)\in\mathfrak{B}$ as exterior differential forms in $z$'s, we endow the dg-algebra $\mathfrak{B}\oplus \mathfrak{M}$ with another differential denoted by $d$.  This is given by the exterior differential on $\mathfrak{B}$ and extends trivially to $\mathfrak{M}$: 
\begin{equation}\label{d}
    da=dz^A\frac{\partial a}{\partial z^A}\,,\qquad dm=0
\end{equation}
for all $a\in \mathfrak{B}$ and $m\in \mathfrak{M}$. It is clear that the operator $d$ has degree $1$, squares to zero,  and differentiates the product (\ref{prod}) in $\mathfrak{B}\oplus \mathfrak{M}$. Furthermore,  the differentials $d$ and $\delta$ commute to each other,  which allows one to combine them into the total differential $D=d+\delta$ of degree $1$.  In such a way we arrive at the dg-algebra $(\mathfrak{B}\oplus\mathfrak{M}, D)$. 

Given the dg-algebra $(\mathfrak{B}\oplus\mathfrak{M}, D)$, one may ask about its cohomology and the minimal model. Since  $\delta$ has form degree $2$, one may regard it as a `small perturbation' of $d$; the differential $d$ increases the form degree only by $1$. Then the standard spectral sequence arguments coupled to  the Poincar\'e lemma for the exterior differential (\ref{d}) show that $H(\mathfrak{B}\oplus\mathfrak{M})\simeq \mathbb{C}[[y]]\oplus \mathbb{C}[[y]]$. (It is isomorphic to the space of $z$-independent elements of $\mathfrak{A}\oplus\mathfrak{M}$, which are obviously nontrivial $d$-cocycles.) The graded space $H(\mathfrak{B}\oplus\mathfrak{M})$ generates the full spectrum of fields -- $\omega$ and $C$ -- in Chiral HiSGRA. 
It turns out that the dg-algebra $\mathfrak{B}\oplus\mathfrak{M}$ is not formal, meaning that the cohomology space 
$H(\mathfrak{B}\oplus \mathfrak{M})$ enjoys higher multiplication operations, in addition to the binary product induced by (\ref{prod}),  making it into a minimal $A_\infty$-algebra. Denoting this $A_\infty$-algebra by $\mathbb{A}$ and tensoring it with the associative algebra $A_1\otimes \mathrm{Mat}_N$, we get a bigger $A_\infty$-algebra   $\mathbb{A}\otimes A_1\otimes \mathrm{Mat}_N$. Applying the standard symmetrization map
 to $\mathbb{A}\otimes A_1\otimes \mathrm{Mat}_N$ gives finally a minimal $L_\infty$-algebra. It is this  ${L}_\infty$-algebra that defines the r.h.s. of field equations (\ref{eq:chiraltheory}). 

Since the unperturbed differential $d$ admits an explicit contracting homotopy $h$ -- that from the Poincar\'e lemma (\ref{homofor}) -- there is a systematic method for constructing the minimal model of the dg-algebra $(\mathfrak{B}\oplus\mathfrak{M}, D)$.  The method results in explicit expressions for the higher multiplication operations in $\mathbb{A}$, see e.g. \cite{merkulov1999strong}, \cite{zbMATH05149113}. For more details we refer the reader to \cite{Sharapov:2018ioy} and to our recent paper \cite{Sharapov:2022faa}, where such a minimal model is constructed for the limiting case $\hhbar=0$.  (All basic steps and formulas are precisely the same.) This provides a rationale for the integrals
and their diagrammatic representation in the main text.

Finally, let us recall that the $A_\infty$-algebra $\mathbb{A}$, being constructed in terms of the partially defined $\star$-product (\ref{sp}), is actually a partial algebra. However, a closer examination \cite[App. C]{Sharapov:2022faa} shows that all the multiplication operations  entering the definition of $\mathbb{A}$ are well defined for the elements of the subspace  $\mathbb{C}[y]\oplus \mathbb{C}[[y]]\subset H(\mathfrak{B}\oplus \mathfrak{M})$. Upon restriction to this subspace we get a total $A_\infty$-algebra that governs the interaction in Chiral HiSGRA.

\section{NNLO vertices}
\label{app:NNLO}
Straightforward evaluation of $G_1$ gives 
\begin{align*}
    G_1&= (p_{12})^2 \exp\Big[\left(1-u_1-u_2\right) p_{01}+\left(1-u_3-u_4\right) p_{02}+u_1 p_{13}+u_2 p_{14}+u_3 p_{23}+u_4 p_{24}+\\
    &\qquad\qquad\qquad\qquad +\hhbar \left(u_4 u_1+u_1+u_2-u_2 u_3-u_3-u_4+1\right)  p_{12}\Big]\,,
\end{align*}
where the original integration variables $t_i$, $i=1,2,3,4$ that are integrated over $[0,1]$ can be identified as 
\begin{align*}
    u_1&=\frac{t_1 \left(t_2-1\right) t_3 t_4}{t_1 t_2 t_3-1}\,,  &  
    u_2&= \frac{t_2 \left(t_1 t_3-1\right) t_4}{t_1 t_2 t_3-1}\,, &
    u_3&= \frac{t_1 \left(t_2 t_3-1\right)}{t_1 t_2 t_3-1}\,, & 
    u_4&= \frac{\left(t_1-1\right) t_2}{t_1 t_2 t_3-1}\,.
\end{align*}
Note that going to $u$'s instead of $t$'s generates a Jacobian that kills a $t$-dependent prefactor, which is the same as in \cite{Sharapov:2022faa}. The corresponding $\mathcal{U}$-vertex is obtained via the duality map:
\begin{align*}
    \mathcal{U}(\omega,C,C,C)&: && \mathcal{U}(p_0,p_1,p_2,p_3,p_4)=\mathcal{V}(-p_4,p_0,p_1,p_2,p_3)\,.
\end{align*}
Similarly, the second tree gives
\begin{align*}
    G_2&=-(p_{13})^2 \exp\Big[\left(1-u_1-u_2\right) p_{01}+\left(1+u_3-u_4\right) p_{03}+u_1 p_{12}+u_2 p_{14}+u_3 p_{23}+u_4 p_{34}+\\
    &\qquad\qquad\qquad\qquad +\hhbar \left(u_4 u_1-u_1+u_2+u_2 u_3+u_3-u_4+1\right)  p_{13}
    \Big]\,,
\end{align*}
where
\begin{align*}
    u_1&=\frac{t_1 \left(t_2 t_3 t_4-1\right)}{t_1 t_2 t_3 t_4-1}\,,  &  
    u_2&= \frac{\left(t_1-1\right) t_2 t_4}{t_1 t_2 t_3 t_4-1}\,, &
    u_3&= -\frac{t_1 \left(t_2-1\right) t_3}{t_1 t_2 t_3 t_4-1}\,, & 
    u_4&= \frac{t_2 \left(t_1 t_3 t_4-1\right)}{t_1 t_2 t_3 t_4-1}\,.
\end{align*}
While not immediately obvious the integrals converge and are over the compact domain in $u$'s. We have also checked directly up to the first few orders that $G_1=\mathcal{V}_1(\omega,\omega,C,C)$ satisfies the 
$L_\infty$-algebra relation
\begin{align*}
    -\omega\star \mathcal{V}_1(\omega,\omega,C,C)+\mathcal{V}_1(\omega\star\omega,\omega,C,C) -\mathcal{V}_1(\omega,\omega\star\omega,C,C)+\mathcal{V}_1(\omega,\omega,\mathcal{U}_1(\omega, C),C)\\
    \qquad+\mathcal{V}_1(\omega,\omega,\mathcal{U}_1(\omega,C,C))-\mathcal{V}_1(\omega,\mathcal{V}_1(\omega,\omega,C),C)=0\,.
\end{align*}
The other $\mathcal{U}$-vertices are obtained exactly as in \cite{Sharapov:2022faa} via the duality map.

\section{NNNLO vertex and beyond}
\label{app:NNNLO}
It is very easy to go to the next level and evaluate a single tree that contributes to the quintic vertex $\mathcal{V}(\omega,\omega,C,C,C)$. The tree grows as

\vspace{-3mm}
$$
   G= \begin{tikzcd}[column sep=small,row sep=small]
       && {} & & &\\
    && \mu\arrow[u] & & &\\
    &  a\arrow[ur]& &\mu\arrow[ul, "h" ']  &&\\
    &&  \mu\arrow[ur,"h"] &  &\arrow[ul]\Lambda[w]&&\\
    & \mu\arrow[ur,"h"] &&\arrow[ul]\Lambda[v]&&& \\
    b\arrow[ur] & & \arrow[ul]\Lambda[u] & &&&
\end{tikzcd}
$$
The answer is a six-fold integral over the `times' $t_{i}$, $i=1,...,6$:
\begin{align*}
    G&= (p_{12})^3\exp\Big[(1-u_1-u_2-u_3) p_{01}+(1-u_4-u_5-u_6) p_{02}+u_1 p_{13}+u_2 p_{14}+\\
    &\qquad\qquad\qquad +u_3 p_{15}+u_4 p_{23}+u_5 p_{24}+u_6 p_{25}+\\
    &\qquad\qquad\qquad +\hhbar\left(-\left(\left(u_3+1\right) \left(u_4+u_5-1\right)\right)-u_6+u_2 \left(1-u_4+u_6\right)+u_1 \left(u_5+u_6+1\right)\right)  p_{12}
    \Big]\,,
\end{align*}
where the integration variables are expressed as
\begin{align*}
    u_1&=\frac{t_1 \left(t_2-1\right) \left(t_3-1\right) t_4 t_5 t_6}{-t_2 t_3 t_5+t_1 t_4 \left(\left(2 t_2-1\right) t_3 t_5-t_2\right)+1} \,, &
    u_2&=\frac{t_2 \left(t_3-1\right) \left(t_1 t_4-1\right) t_5 t_6}{-t_2 t_3 t_5+t_1 t_4 \left(\left(2 t_2-1\right) t_3 t_5-t_2\right)+1} \,,\\
    u_3&= \frac{t_3 \left(-t_2 t_5+t_1 t_4 \left(t_2 \left(2 t_5-1\right)-t_5\right)+1\right) t_6}{-t_2 t_3 t_5+t_1 t_4 \left(\left(2 t_2-1\right) t_3 t_5-t_2\right)+1}\,, &
    u_4&=\frac{t_1 \left(-t_2 t_4+t_3 \left(t_2 \left(2 t_4-1\right)-t_4\right) t_5+1\right)}{-t_2 t_3 t_5+t_1 t_4 \left(\left(2 t_2-1\right) t_3 t_5-t_2\right)+1} \,,\\
    u_5&= \frac{\left(t_1-1\right) t_2 \left(t_3 t_5-1\right)}{-t_2 t_3 t_5+t_1 t_4 \left(\left(2 t_2-1\right) t_3 t_5-t_2\right)+1}\,,&
    u_6&= \frac{\left(t_1-1\right) \left(t_2-1\right) t_3}{-t_2 t_3 t_5+t_1 t_4 \left(\left(2 t_2-1\right) t_3 t_5-t_2\right)+1}\,.
\end{align*}
It is again pleasing how local the vertex is: there are no $p_{34}$, $p_{45}$ and $p_{35}$. The corresponding $\mathcal{U}$-vertex is obtained via the duality map:
\begin{align*}
    \mathcal{U}(\omega,C,C,C,C)&: && \mathcal{U}_1(p_0,p_1,p_2,p_3,p_4,p_5)=\mathcal{V}_1(-p_5,p_0,p_1,p_2,p_3,p_4)\,.
\end{align*}
It is easy to iterate the tree and find the all order result for $\mathcal{V}(\omega,\omega,C,...,C)$:
\begin{align*}
    G&= (p_{12})^n \exp\Big[ (1-\sum_i u_i) p_{01} +(1-\sum_i v_i) p_{02} +\sum_i u_i p_{1,i+2}+\sum_i v_i p_{2,i+2}+ \\
     &\qquad \qquad \qquad \qquad +\hhbar\, \Big(1+\sum_i (u_i-v_i) +\sum_{i,j} u_iv_j \sign(j-i) \Big ) p_{12} \Big]\,.
\end{align*}
Here all sums are from $1$ to $n$ and $n$ is the number of $C$-fields. By definition, we set $\sign(0) =0$. This also gives $\mathcal{U}(\omega,C,C,...,C)$ via the duality map. Other trees lead to similar expressions.

\footnotesize
\providecommand{\href}[2]{#2}\begingroup\raggedright\endgroup

\end{document}